\documentclass[dvips]{article}

\usepackage{icrc2011}

\title{A Multi-Wavelength Study of the Unidentified TeV $\gamma$-ray Source \hj}

\newcommand{\etal}{\MakeLowercase{\textit{et al. }}} 
\newcommand{\ergcm}[1]{$\times 10^{#1}$ erg cm$^{-2}$ s$^{-1}$}
\newcommand{\ergcmalt}[1]{$10^{#1}$ erg cm$^{-2}$ s$^{-1}$}

\newcommand{\nh}{N$_{\rm H}$}

\newcommand{\HI}{\ion{H}{I}}

\newcommand{\ltsima}{$\buildrel < \over \sim$}
\newcommand{\lsim}{\lower.5ex\hbox{\ltsima}}
\newcommand{\gtsima}{$\buildrel > \over \sim$}
\newcommand{\gsim}{\lower.5ex\hbox{\gtsima}}

\newcommand{\xmm}{\emph{XMM-Newton}}
\newcommand{\hess}{H.E.S.S.}
\newcommand{\nanten}{\emph{Nanten}}

\newcommand{\hj}{HESS\,J1626$-$490}
\newcommand{\snr}{SNR\,G335.2+00.1}

\newcommand{\lmxb}{4U\,1624-490}

\newcommand{\arcmin}{$^\prime$}
\newcommand{\arcsec}{$^{\prime\prime}$}
\newcommand{\apj}{ApJ}

\newcommand{\memsai}{Memorie della Societa Astronomica Italiana}

\newcommand{\aap}{A{\&}A}

\newcommand{\mnras}{MNRAS}

\newcommand{\aj}{AJ}
\newcommand{\araa}{ARAA}

\newcommand\ion[2]{#1$\,${\scshape{#2}}}

\shorttitle{Eger \etal multi-wavelength study of \hj}

\authors{P.\,Eger$^{1}$, G.\,Rowell$^{2}$, A.\,Kawamura$^{3}$, Y.\,Fukui$^{3}$, L.\,Rolland$^{4}$, C.\,Stegmann$^{1}$}
		
\afiliations{
$^1$Erlangen Centre for Astroparticle Physics, Universit\"at Erlangen-N\"urnberg, Erwin-Rommel-Str. 1, 91058 Erlangen, Germany\\
$^2$School of Chemistry and Physics, University of Adelaide, Adelaide 5005, Australia\\
$^3$Department of Astrophysics, Nagoya University, Furocho, Chikusaku, Nagoya 464-8602, Japan\\
$^4$CEA Saclay, DSM/IRFU, F-91191 Gif-Sur-Yvette Cedex, France; now at Laboratoire d'Annecy-le-Vieux de Physique des Particules (LAPP), Universit\'e de Savoie, CNRS/IN2P3, F-74941 Annecy-Le-Vieux, France\\
}

\email{peter.eger@physik.uni-erlangen.de}

\abstract{
\hj , so far only detected with the H.E.S.S. array of imaging atmospheric 
Cherenkov telescopes, could not be unambiguously identified with any source seen at
lower energies. 

Therefore, we analyzed data from an archival \xmm\ observation, pointed towards
\hj , to classify detected point-like and extended X-ray sources according 
to their spectral properties. 
None of the detected X-ray point sources fulfills the energetic
requirements to be considered as the synchrotron radiation (SR) counterpart to the
VHE source assuming an Inverse Compton (IC) emission scenario. Furthermore, we did
not detect any diffuse X-ray excess emission originating from the region around \hj\ 
above the Galactic Background. 
The derived upper limit for the total X-ray flux disfavors a purely leptonic emission scenario for \hj . 

To characterize the Interstellar Medium surrounding \hj\ we analyzed
$^{12}$CO(J=1-0) molecular line data from the \nanten\ Galactic plane survey and \HI\ data
from the Southern Galactic Plane Survey (SGPS). 
We found a good morphological match between molecular and atomic gas
in the $-27$~km/s to $-18$~km/s line-of-sight velocity range and \hj . 
The cloud has a mass of 1.8$\times^4$M$_\odot$ and is located at a mean kinematic distance of
$d$ = 1.8~kpc. Furthermore, we found a density depression in the \HI\ gas at a similar
distance which is spatially consistent with the SNR G335.2+00.1. 

Therefore, the most likely origin of the VHE $\gamma$-ray emission observed with
\hess\ is the hadronic interaction of cosmic rays with a moderately dense molecular
cloud, which we detected with \nanten . 
The application of a detailed hadronic model for cosmic ray transport and interaction shows 
that the cosmic rays could originate from the nearby \snr .
}

\begin{document}
\maketitle

\section{Introduction}
\hj\ is a VHE $\gamma$-ray source of unknown origin, which so far could not be identified with any counterpart 
at lower wavelengths. 
This object, with an intrinsic extension of $\sim$5~arcmin (Gaussian FWHM), is located right on the Galactic plane 
(R.A.: 16$^h$26$^m$04$^s$, Dec.: $-49^\circ$05\arcmin13\arcsec) and was detected by \hess\ with a peak significance 
of 7.5$\sigma$ \cite{2008A&A...477..353A}. 
These authors measured a power-law spectrum with a photon index of 
2.2~$\pm$~0.1$_{stat}$~$\pm$~0.2$_{sys}$ and a flux normalization of 
\hbox{(4.9$\pm$0.9)$\times$10$^{-12}$cm$^{-2}$\,s$^{-1}$\,TeV$^{-1}$} at 1\,TeV at energies between 0.5~TeV and 40~TeV. 

In this work we analyzed the data of an archival \xmm\ observation to search for an X-ray 
counterpart to \hj .
Therefore, we classified the detected X-ray point sources in the vicinity of \hj\ and searched for possible diffuse 
excess emissions above the expected Galactic background. 
Furthermore, we present $^{12}$CO(J=1-0) molecular line survey data taken with the \nanten\ 
mm/sub-mm observatory to scan for molecular clouds, and SGPS \HI\ data to characterize the neutral interstellar medium.

\section{\xmm\ X-ray data analysis and results}
We analyzed the EPIC-pn and MOS data of an archival \xmm\ dataset (ID: 0403280201) pointed towards the direction of \hj .
After screening for periods of enhanced background activity the net exposures for pn and MOS are 
4.9~ks and 13.2~ks, respectively. 

Using the standard maximum likelihood technique of the Science Analysis Software (SAS) we searched for 
X-ray point sources within the 4$\sigma$ VHE $\gamma$-ray contours of \hj . 
Twelve such sources were detected in this region above a limiting flux of $\sim$2\ergcm{-14}. 
We characterized these sources according to their X-ray spectral properties as well as their 
near-infrared counterpart from the 2MASS all-sky catalog \cite{2006AJ....131.1163S}. 
Based on this classification scheme (for more details see \cite{2011A&A...526A..82E}) 
we calculated the unabsorbed fluxes for all detected point sources, 
assuming different spectral shapes and foreground hydrogen column densities, depending on the source class. 
By far the brightest detected X-ray source features a soft thermal X-ray spectrum with an unabsorbed flux of 7.9\ergcm{-12}
(0.5 -- 10~keV), and can be identified with a triplet system of coronally active main sequence stars. 
All other X-ray sources feature unabsorbed fluxes $<$3.5\ergcm{-13} (0.5 -- 10~keV). 

 \begin{figure}[!t]
  \centering
  \includegraphics[width=0.98\hsize]{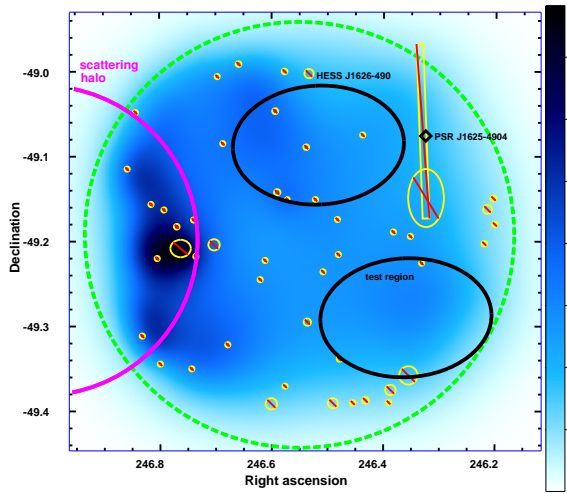}
  \caption{Adaptively smoothed EPIC-PN image of diffuse X-ray flux in the 
  		   3--7~keV energy band with a linear color scale (arbitrary units). 
		   The instrument FoV is shown as a dashed circle (green). 
		   The large ellipses (black) denote the extraction regions for \hj\ and 
		   the background test, respectively. 
		   The large circle (magenta) to the east gives the approximate extension of the X-ray scattering halo 
		   of \lmxb . 
		   Excluded regions are shown as crossed-out areas (yellow regions crossed with red lines)}
  \label{fig-x-ray-diffuse}
 \end{figure}

We also searched for a potential extended and diffuse X-ray component related to \hj . 
\cite{2008A&A...477..353A} approximated the morphology of \hj\ by fitting a 2-D Gaussian to the 
VHE $\gamma$-ray excess map.
The resulting intrinsic (with the effects of the instrument point-spread-function removed) major and minor 
axes are 0.1$^\circ$ and 0.07$^\circ$, respectively, with a position angle of 3$^\circ$ west to north. 
From the same region we searched for diffuse and extended X-ray emission above the astrophysical and instrumental 
background components.
Figure~\ref{fig-x-ray-diffuse} shows a smoothed image of the diffuse X-ray flux in the 3--10~keV band. 
Point-like sources were removed, based on the 99\% energy containment radius of the PSF. 
Also, the read-out streak caused by the brightest source in the FoV has been excluded. 
The instrumental and particle-induced background components were subtracted using a non-X-ray background dataset 
from the calibration database. 
Apart from some contamination towards the east of the FoV 
due to an X-ray scattering halo caused by the bright binary system \lmxb , the level of diffuse X-ray emission
appears to be homogeneous and flat throughout the FoV. 
To quantify potential diffuse excess X-ray emission from the direction of \hj\ and to compare the signal to the 
expected diffuse Galactic background, we extracted a spectrum from an elliptical region compatible with the 
intrinsic VHE $\gamma$-ray extent, as well as from a test region in the southern part of the \xmm\ FoV 
(as indicated in Fig.~\ref{fig-x-ray-diffuse}). 
We found no indication for enhanced X-ray emission from \hj\ compared to the test region, and the spectra 
from both regions are compatible with typical diffuse Galactic X-ray emission expected from such a region 
in the Galactic plane (see, e.g., \cite{2005ApJ...635..214E}). 
We derived an upper limit for extended X-ray emission connected to \hj\ of $F_\mathrm{X,excess} < 4.9$\ergcm{-12} 
(1--10~keV), assuming a powerlaw spectral shape with index $-2$.

\section{\nanten\ $^{12}$CO(J=1-0) and SGPS \HI\ data}
\label{sec-co-hi}
To search for molecular clouds, spatially and morphologically coincident with \hj , 
we analyzed $^{12}$CO(J=1-0) molecular line observations performed by the 4~m mm/sub-mm 
\nanten\ telescope, located at Las Campanas Observatory, Chile \cite{2004ASPC..317...59M}. 
For this work the local standard of rest velocity ($v_{\rm LSR}$) range $-240$ to +100 km\,s$^{-1}$ was searched. 

 \begin{figure}[!t]
  \centering
  \includegraphics[width=0.98\hsize]{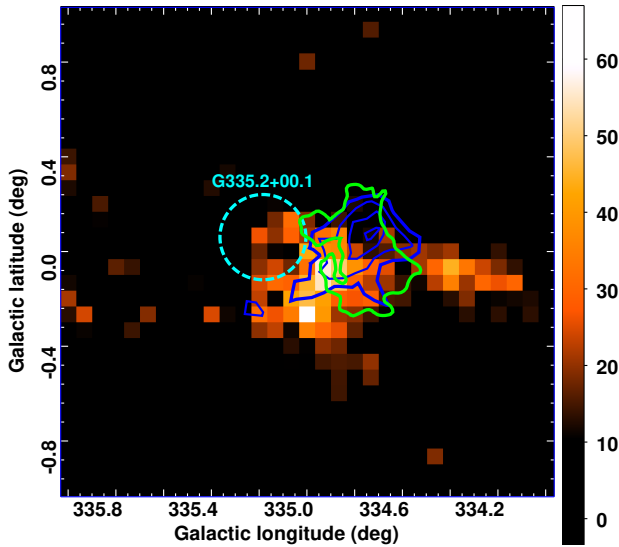}
  \caption{\nanten\ $^{12}$CO(J=1-0) image of the region around \hj\ (linear scale in K\,km\,s$^{-1}$) 
  		integrated over the $v_{\rm LSR}$ range $-31$ to $-18$~km\,s$^{-1}$. 
		Overlaid are the contours of the VHE emission (blue) and of the \HI\ cloud discussed in Sect.~\ref{sec-co-hi} 
		(green). 
		The dashed circle (light blue) denotes the position and extension of \snr .}
  \label{fig-co-image}
 \end{figure}

Figure~\ref{fig-co-image} shows the \nanten\ $^{12}$CO(J=1-0) image integrated over 
the $v_{\rm LSR}$ range $-31$ to $-18$~km\,s$^{-1}$. 
In this interval we found a $^{12}$CO feature partially overlapping with the VHE emission. 
According to the Galactic rotational model of \cite{1993A&A...275...67B}, this $v_{\rm LSR}$ range corresponds 
to a kinematic distance of 1.5 to 2.2~kpc. 
Using the relation between the hydrogen column density (N$_{\rm H}$) and the $^{12}$CO(J=1-0) 
intensity W($^{12}$CO), \hbox{\nh\ = 1.5 $\times$ 10$^{20}$ [W($^{12}$CO)/K\,km/s]} \cite{2004A&A...422L..47S}, 
we estimate the total mass of this cloud at 1.8$\times 10^4$M$_{\odot}$ for $d$ = 1.8~kpc 
within an elliptical region centered at $l$ = 334.78, $b$ = 0.00 with dimensions 0.26$\times$0.30~deg.
The corresponding average density is 2.1$\times$10$^2$~cm$^{-3}$. 

\begin{figure}[!t]
  \centering
  \includegraphics[width=0.98\hsize]{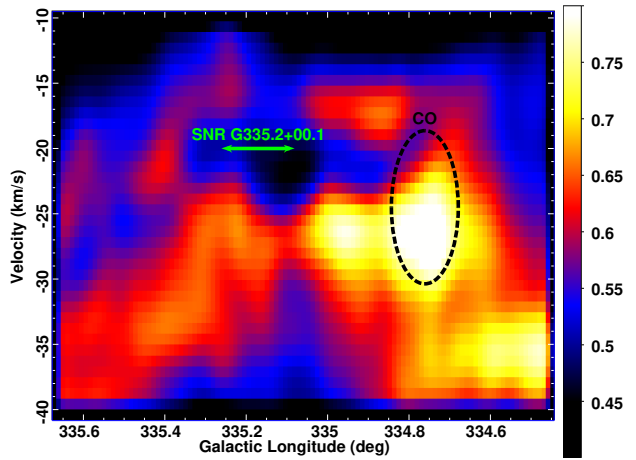}
  \caption{\HI\ SGPS Galactic longitude--velocity plot (linear scale in K\,deg) 
			integrated over the Galactic latitude range $-0.11$ to 0.24~deg. 
			The position and extension of the $^{12}$CO cloud is denoted by a dashed circle (black). 
			The double arrow (green) shows the Galactic latitude and extension of \snr .}
  \label{fig-hi-pvplot}
 \end{figure}

In addition to the \nanten\ $^{12}$CO molecular line data which trace the densest regions of the interstellar gas, we 
also investigated the environment of \hj\ using \HI\ data from the Southern Galactic Plane Survey (SGPS). 
Figure~\ref{fig-hi-pvplot} shows the $v_{\rm LSR}$ profile for this region integrated over the Galactic 
latitude range $-0.11$ to 0.24~deg, which is also the extent of the VHE $\gamma$-ray signal of \hj . 
This image shows a region of increased gas density in spatial coincidence with the CO molecular cloud, 
most pronounced in the $-31$ to $-23$~km\,s$^{-1}$ velocity range. 
The contours extracted from an \HI\ image integrated over this velocity range are shown in Fig.~\ref{fig-co-image}. 
We estimated the mass of the dense \HI\ region coinciding with the VHE and $^{12}$CO features using the signal 
within the same elliptical region as for the CO cloud (Sect.~\ref{sec-co-hi}). 
Using the relation between \HI\ intensity and column density from \cite{1990ARA&A..28..215D} 
(X=1.8$\times 10^{18}$\,cm$^{-2}$\,K$^{-1}$\,km/s) we estimated the mass of the cloud as 
4.9$\times 10^{3}$\,${\rm M_{\odot}}$ with an average density of 60.1~cm$^{-3}$. 
Furthermore, a local \HI\ density depression is seen in the center of the image at an angular separation 
of $\sim$21\arcmin\ from the CO molecular cloud and \hj . 
It is striking that this feature is consistent in position as well as in angular extension with the \snr\ 
\cite{2009BASI...37...45G,2006AJ....131.1479R}. 

\section{Discussion}
In this section we discuss potential emission scenarios giving rise to the VHE $\gamma$-ray signal detected 
with \hess , particularly in the context of the presented multi-wavelength data. 

In the case of a leptonic scenario where low-energy photons are up-scattered by relativistic electrons 
via the IC process, X-ray emission is expected to accompany the VHE $\gamma$-ray signal arising from synchrotron cooling of the 
same population of high-energy electrons \cite{1997MNRAS.291..162A}. 
Even from a first glance at the flux levels in X-rays ($\sim$\ergcmalt{-13} for point sources and $\sim$\ergcmalt{-12} for 
diffuse emission) and at VHE $\gamma$-rays ($\sim$\ergcmalt{-11}), a purely leptonic emission scenario seems unlikely. 
Following \cite{1997MNRAS.291..162A} and assuming a typical Galactic magnetic field strength of 3$\times$10$^{-6}$~G, 
and a photon index of $\Gamma$ = 2.2 (as in the VHE domain), 
we estimate an integrated X-ray source flux of $\sim$1.1\ergcm{-11} (0.5--10.0~keV). 
This flux is a factor of $\sim$25 more than what we measured from all but one point source 
which makes an identification of \hj\ with any of these sources unlikely. 
As the only sufficiently bright X-ray point source in the FoV can be identified with a triplet system of active main sequence 
stars it is most likely not related to \hj . 
Furthermore, the upper limit for diffuse X-ray emission is a factor of $\sim$2 lower 
than the expected value for \hj\ in a purely leptonic model. 

Not detecting any X-ray source fulfilling the energetic requirements for a purely leptonic 
scenario favors a hadronic emission process such as dense clouds in the vicinity of a cosmic particle accelerator. 
Such a scenario will be discussed in the remaining part of this section. 
Dense molecular clouds are established VHE $\gamma$-ray emitters because 
they provide target material in regions of high cosmic-ray densities. 
Using data from the \nanten\ $^{12}$CO(J=1-0) Galactic plane survey, we detected a 
molecular cloud that is morphologically consistent with \hj . 
This object is located at a kinematic distance of $\sim$1.8~kpc. 
Following \cite{1994A&A...285..645A} (Eqs. 2 and 3), we estimated the required gas density to produce 
the observed VHE $\gamma$-ray signal (F$_\gamma$($>$0.6\,TeV) = 7.5$\times$10$^{-12}$~ph~cm$^{-2}$~s$^{-1}$ and 
$\Gamma$ = 2.2) as $n$ $\approx$ 10 cm$^{-3}$ assuming a cosmic-ray production efficiency 
of $\theta$ = 0.1 and a distance of $d = 1.8$~kpc.
This value is an order of magnitude lower than the measured $^{12}$CO mean density. 
Thus, this environment would be easily suited to providing the observed VHE $\gamma$-ray flux. 

Now assuming that this $^{12}$CO / \HI\ cloud is indeed the source of the observed VHE $\gamma$-rays, 
a cosmic-ray accelerator would be needed in its vicinity. 
The nearby density depression seen in \HI\ (see Sect.~\ref{sec-co-hi}) might indicate the presence of a 
recent catastrophic event, such as an SNR, giving rise to strong shocks that would have blown the neutral gas out. 
At $d = 1.8$~kpc the edge of this region would be at a distance of 8.1~pc from the $^{12}$CO/\HI\ cloud. 
It is striking that this \HI\ feature is consistent in both position and angular extension with the \snr\ 
\cite{2009BASI...37...45G}. 
Using the $\Sigma$-D relation the distance to \snr\ was estimated as $d = 5$~kpc \cite{2004SerAJ.169...65G}, 
which conflicts with the identification of \snr\ with the \HI\ depression seen at a distance of 1.8~kpc.
However, the scatter in $\Sigma$-D distances for individual sources is quite large (see \cite{2005MmSAI..76..534G}) 
making a positive identification of \snr\ and the \HI\ depression seen at a kinematic distance of 1.8~kpc very well possible. 

 \begin{figure}[!t]
  \centering
  \includegraphics[width=0.7\hsize,angle=-90,clip]{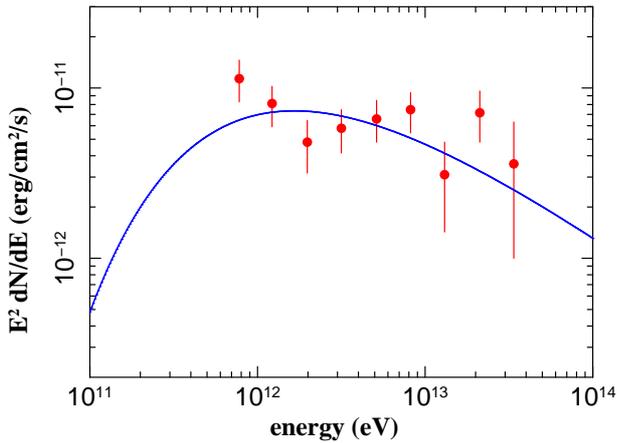}
  \caption{Spectral energy distribution of \hj\ showing the VHE $\gamma$-ray spectrum measured with \hess\ (red). 
	The solid line (blue) shows the hadronic model with the parameters that yield the best fit 
	to the data	($R = 35$~pc, $t = 1.5\times 10^5$~yr, $D_{10} = 9\times 10^{25}$cm$^2$~s$^{-1}$).}
  \label{fig-sed}
 \end{figure}

To explore the physical scenario of the interaction of the \snr\ with the $^{12}$CO molecular cloud in more detail, 
we adopted the model described by \cite{2009MNRAS.396.1629G}. 
In this model hadronic cosmic rays with a powerlaw energy spectrum with index $-2$ and a total energy 
of $10^{50}$~erg are injected at the SNR shock into the ISM. 
The transport of these cosmic rays is described by an energy-dependent diffusion coefficient of the form 
$D_\mathrm{ISM} = D_{10}\cdot E^{0.5}$, where $E$ is the energy of the cosmic rays and $D_{10}$ is the diffusion 
coefficient for an energy of 10~GeV. 
For the assumptions made by \cite{2009MNRAS.396.1629G}, the transport equation can be solved analytically and 
the expected $\gamma$-ray spectrum due to the production and subsequent decay of neutral pions can be calculated. 
Some of the model parameters can be constrained by the measurements presented in this work. 
These are the mass $M$, the density $n$ of the molecular cloud and the distance $d$ between the object and the observer. 
Also, a lower limit for the physical distance $R$ between the SNR shock and the cloud was derived, which corresponds 
to the projected distance between these two objects (8.1~pc). 
However, it is still possible that the actual separation between SNR and cloud is in fact larger than this value due 
to slightly different line-of-sight distances between these two objects and the observer. 
The remaining free model parameters are the age of the SNR $t$, the diffusion coefficient $D_{10}$ and the 
radius $R$. 
Figure~\ref{fig-sed} shows the VHE $\gamma$-ray spectrum of \hj\ measured with \hess , together with 
the model with the parameters that yielded the best fit to the data 
($R = 35$~pc, $t = 1.5\times 10^5$~yr, $D_{10} = 9\times 10^{25}$cm$^2$~s$^{-1}$). 
The best-fit value of $D_{10}$ is about two orders of magnitude lower than the estimated mean value for the 
Galactic plane, however, it agrees well with comparable measurements of diffusion in the vicinity of other 
cosmic-ray accelerators (e.g., for the W28 region, see \cite{2010sf2a.conf..313G}). 
It has to be noted here, however, that the age $t$ and the diffusion coefficient $D_{10}$ are strongly 
correlated, and as $t$ is not known for \snr\ $D_{10}$ is not very well constrained. 
Overall though, the modeling shows that a hadronic scenario is indeed viable for \hj\ with reasonable 
physical parameters. 
Another complication of hadronic models is the diffusion of cosmic rays into the target material which strongly depends on the 
density profile of the molecular cloud. 
More detailed and better resolved molecular line observations could provide the necessary data in the future.

\clearpage


\begin{thebibliography}{10}

\bibitem{2008A&A...477..353A}
{Aharonian}, F., et~al.
\newblock \aap, \textbf{477} (2008): 353--363

\bibitem{2006AJ....131.1163S}
{Skrutskie}, M.~F., et~al.
\newblock \aj, \textbf{131} (2006): 1163--1183

\bibitem{2011A&A...526A..82E}
{Eger}, P., et~al.
\newblock \aap, \textbf{526} (2011): A82+

\bibitem{2005ApJ...635..214E}
{Ebisawa}, K., et~al.
\newblock \apj, \textbf{635} (2005): 214--242

\bibitem{2004ASPC..317...59M}
{Mizuno}, A., {Fukui}, Y.
\newblock {D.~Clemens, R.~Shah, \& T.~Brainerd}, editor, Milky Way Surveys: The
  Structure and Evolution of our Galaxy, vol. 317 of \emph{Astronomical Society
  of the Pacific Conference Series}. 2004 59--+

\bibitem{1993A&A...275...67B}
{Brand}, J., {Blitz}, L.
\newblock \aap, \textbf{275} (1993): 67--+

\bibitem{2004A&A...422L..47S}
{Strong}, A.~W., et~al.
\newblock \aap, \textbf{422} (2004): L47--L50

\bibitem{1990ARA&A..28..215D}
{Dickey}, J.~M., {Lockman}, F.~J.
\newblock \araa, \textbf{28} (1990): 215--261

\bibitem{2009BASI...37...45G}
{Green}, D.~A.
\newblock Bulletin of the Astronomical Society of India, \textbf{37} (2009):
  45--+

\bibitem{2006AJ....131.1479R}
{Reach}, W.~T., et~al.
\newblock \aj, \textbf{131} (2006): 1479--1500

\bibitem{1997MNRAS.291..162A}
{Aharonian}, F.~A., {Atoyan}, A.~M., {Kifune}, T.
\newblock \mnras, \textbf{291} (1997): 162--176

\bibitem{1994A&A...285..645A}
{Aharonian}, F.~A., {Drury}, L.~O., {Voelk}, H.~J.
\newblock \aap, \textbf{285} (1994): 645--647

\bibitem{2004SerAJ.169...65G}
{Guseinov}, O.~H., {Ankay}, A., {Tagieva}, S.~O.
\newblock Serbian Astronomical Journal, \textbf{169} (2004): 65--+

\bibitem{2005MmSAI..76..534G}
{Green}, D.~A.
\newblock \memsai, \textbf{76} (2005): 534--+

\bibitem{2009MNRAS.396.1629G}
{Gabici}, S., {Aharonian}, F.~A., {Casanova}, S.
\newblock \mnras, \textbf{396} (2009): 1629--1639

\bibitem{2010sf2a.conf..313G}
{Gabici}, S., et~al.
\newblock {S.~Boissier, M.~Heydari-Malayeri, R.~Samadi, \& D.~Valls-Gabaud },
  editor, Proceedings of the Annual meeting of the French Society of Astronomy
  and Astrophysics. 2010 313--+

\end{thebibliography}
\end{document}